\documentclass[preprint,12pt,3p,times,sort&compress]{elsarticle}
\usepackage{graphicx}
\usepackage[T1]{fontenc}
\usepackage[utf8]{inputenc}
\usepackage{mathtools,url}

\newlength{\figwidthlarge} 			
\newlength{\figwidth} 			
\newlength{\figwidthsmall} 			
\usepackage[usenames,dvipsnames]{color}
\definecolor{gray}{rgb}{0.7,0.7,0.7}
\definecolor{dblue}{rgb}{0,0,0.75}
\definecolor{dred}{rgb}{0.6,0,0}
\definecolor{dgreen}{rgb}{0,0.5,0}

\journal{Physica A}

\begin{document}

\begin{frontmatter}

\title{Impact of independence on polarization of opinions}

\author[WMat]{Janusz Szwabi\'nski\corref{cor1}}
\ead{janusz.szwabinski@pwr.edu.pl}
\author[WMat]{Tomasz Weron}
\ead{tomek.weron@gmail.com}

\address[WMat]{Faculty of Pure and Applied Mathematics, Wroc{\l}aw University of Science and Technology, 50-370 Wroc{\l}aw, Poland}

\cortext[cor1]{Janusz Szwabi\'nski}

\date{This version: \today}

\begin{abstract}
Polarization of societies is getting more and more attention from researchers working at the intersection of many fields, because it seems to be a defining feature of many public domains. In this paper, we are going to investigate how the unwillingness to yield to the group pressure, also known as independence, influences this phenomenon. In particular, we would like to answer the question whether independent choices of people could alter the dynamics of a system that otherwise would become polarized. A modified version of the $q$-voter model will be used for that purpose. From our findings it follows that the impact of independence is at least two-fold. At low independence levels the consensus-polarization transition between two antagonistic groups sets in quicker than in the absence of independence. Higher levels induce additional transition in the system, from a polarized state to a disordered one.
\end{abstract}


\end{frontmatter}

\section{Introduction}
\label{sec:Introduction}

Polarization is a concept used frequently in social and political science as well as economics, but its definition may differ between domains.  Within this paper we will follow the one given by DiMaggio et al and assume that polarization refers to a situation in which a group of people is divided into two opposing cliques having contrasting positions on a given issue~\cite{DIM96}. This type of polarization is sometimes called bi-polarization~\cite{MAS13} to distinguish it from the group polarization phenomenon, i.e. the tendency for a group to make decisions more extreme than the initial inclination of its members~\cite{ISE86,SUN02}.

Recent observers point to a growing polarization of modern societies~\cite{BAI18}. It seems to be a defining feature of many public domains and was identified in the World Economic Forum's 2017 Global Risk Report already as one of the top threats to the global order~\cite{GLO17}. Consequently, it is getting more and more attention from researchers working at the intersection of many fields, including social and political science, economics, mathematics and statistical physics.

High and increasing levels of polarization are attributed to a variety of sources, including the isolating effects of social media or news outlets focusing rather on outraged rants than reasoned debates. Although a significant progress in our understanding of polarization mechanisms has been observed in recent years, our knowledge remains sketchy and there is still a lot of room for improvement. And every new insight into polarization is important, because it is known to have a huge impact on societies. It leads to social tensions and conflicts, and often it may end up in segregation of societies~\cite{DIM96}. 

Interestingly, not all debates have the potential to polarize societies. From the observations it follows that in order to drive people to extreme and opposing opinions, the topic of a discussion has to be perceived as important by all participants and emotionally charged. That is why polarizing topics comprise controversial issues such as abortion rights, homosexuality, public funding for the art, gun control, global warming, vaccination and last but not least - politics~\cite{MOU01,MCC11,GRU14,ADA05,MAO06,WAU11}.

Starting with Eli Pariser's book~\cite{PAR11}, social media sites are more and more blamed for intensifying (political) polarization. The artificial intelligence algorithms used by sites such as Facebook, Twitter or Google to profile the users create so called ``echo chambers'' (or  ``filter bubbles'') which separate people from the information that disagrees with their viewpoints. The idea behind those algorithms was to let the people stay in their comfort zones. An unexpected side effect of this approach is an unconscious confirmation bias, because people are confronted mainly with information which reinforces their beliefs and opinions. 
The bias may contribute to overconfidence in personal beliefs and can maintain or strengthen them in the face of contrary evidence, which leads to polarization~\cite{BAI18}.

Several possible mechanisms leading to a stable bi-polar distribution of opinions within a simulation have been already proposed. There is for instance a series of papers showing that \textit{opinion homophily} may support opinion plurality including polarization~\cite{AXE97, MCP01, HEN02, DEF00}. This type of homophily is understood as a relationship between similarity of peoples' views and an increased likeliness of their interactions. It was usually implemented as a \textit{bounded confidence}, i.e. a threshold mechanisms that switches off influence in case the opinion discrepancy is too big. Long range ties (bridges between clusters) in a social network may also foster polarization if homophily and assimilation at the microlevel are combined with some negative influence, e.g. xenophobia~\cite{MAC03,SAL06}. From social balance theories it follows that a mixture of positive and negative ties is needed for polarization to emerge and to prevail~\cite{TRA13,SIE16,KRU17}. In the argument-communication model agents with a similar attitude mutually reinforce that attitude by the exchange of supportive arguments, which in some circumstances leads to polarization as well~\cite{MAS13}. And finally inflexibility, understood here as an internal opinion that encodes how many encounters of different-minded agents are needed for an agent to change its external opinion, has been shown to polarize a population in the sense that two opposing camps of more and more inflexible supporters may emerge~\cite{MAR08,MAR13,BAN19}.

Recently, we proposed a simple model of polarization based on the $q$-voter model with both conformity and anticonformity~\cite{SIE16,KRU17}. We considered the model on a double-clique social network, because it mimics the echo chambers observed on social media platforms as well as the interactions between their members. We found that if the number of inter-clique connections stays below a critical value, a consensus between two antagonistic cliques is possible. Thus, in light of these results the artificial intelligence algorithms producing echo chambers on many platforms may have a positive impact in terms of polarization, because they are reducing the exposure to different opinions. In this paper we are going to extend our model with independence to make the spectrum of possible responses to social influence more realistic from the social science perspective~\cite{NAI86,NAI00,NYC13}.

The paper is organized as follows. In the next section we introduce two versions of our model -- a quenched and an annealed one. Then, we will investigate the model both analytically and numerically. Finally, some conclusions will be presented.

\section{Model}
\label{sec:Model}

\subsection{Basic assumptions}

The basic assumptions of the model have been already extensively discussed in Refs.~\cite{SIE16,KRU17}. Therefore we start this section with only a short overview of its major premises:
\begin{itemize}
\item a binary opinion model with a single trait,
\item $q$-voter model with conformity and anticonformity as the general modeling framework,
\item double clique topology of the social network,
\item conformity between agents within a clique and anticonformity in the interactions between the cliques.
\end{itemize}
It should be emphasized that all of the above assumptions can be justified by recent findings in the opinion dynamics community. For instance, the analysis of many social networks revealed that the polarization of opinions within those networks may be correlated with their segmentation~\cite{CON11,NEW06,ZAC77}. Hence, we assumed that the network is already modular and took the double clique topology~\cite{SOO08} as its model. Choosing a binary model with a single trait roots in the observation that in many situations opinions of people may be interpreted as simple ``yes/no'' (i.e. binary) answers~\cite{WAT07}. Moreover, social networks are often characterized by a semantic unicity, i.e. opinions and interactions of networks' members are restricted to a single topic~\cite{GUE13}.

The $q$-voter model is one of the extensively studied models of binary opinions. Within the original formulation~\cite{CAS09b}, $q$ randomly picked neighbors influence a voter to change his opinion. The voter conform if all $q$ neighbors agree. Otherwise he is allowed to flip his opinion with a probability $\epsilon$. This unanimity rule embedded in the model  is in line with a number of social experiments~\cite{MYE13,BON05,ASC55}.

Conformity, understood as the act of matching opinions to the group norm, is the only social force in the original $q$-voter model. However, it is relatively easy to extend it with other possible responses to social influence like independence and/or anticonformity~\cite{NYC12,NYC13,NYC18,PRZ14,SZN14b,JED16,JED17,JED18,JED18b}. The first one is simply the unwillingness to yield to the group pressure and introduces noise to the system, the latter means a deliberate challenging the position of the group. In Refs.~\cite{SIE16,KRU17} we used anticonformity to mimic negative ties between agents belonging to two opposite cliques, in agreement with the social balance theories~\cite{TRA13}. It should be noted that the double clique topology with conformity inside a clique and anticonformity between the cliques resembles to some extent the controversial echo chambers generated by social platforms~\cite{PAR11}.

\subsection{Independence of agents}

In Refs.~\cite{SIE16,KRU17} we have shown, both theoretically and by means of Monte Carlo simulations, that a system consisting of two antagonistic cliques connected with each other undergoes a phase transition as the number of cross-links between the cliques changes. Below the critical point (i.e. loosely connected cliques) the intra-clique conformity takes over and consensus in the entire system as an asymptotic state is possible. Above the  critical point the system ends up in a polarized state with the cliques having opposite opinions and local consensus within each of them. It was a surprising result, because it actually defies the criticism of echo chambers started by Pariser~\cite{PAR11}. Since the algorithms generating the echo chambers reduce the exposure time to different-minded people, in the  light of our findings they should lower the polarization level between antagonistic groups instead of enhancing it. 

However, one of the drawbacks of the model presented in Refs.~\cite{SIE16,KRU17} was the lack of independence in the behavior of agents. This concept has been already considered in a series of models~\cite{GAL91,GAL97,GAL07,MOB03,NYC12,NYC13}. It actually implies the failure of an attempted social influence, because an independent individual makes decisions independently of the group norm. From the perspective of social science it falls (together with anticonformity) into the category of non-conformal behaviors~\cite{NAI86,NAI00,NAI13}. From a physical point of view it plays the role of social temperature that induces an order-disorder transition~\cite{GAL04,SZN11,NYC12}. Thus, it would be interesting to check how the introduction of independence into our model will change the behavior of the entire system and if our findings will still hold in the extended version of the model.

\begin{figure}
\centering
\includegraphics[scale=0.7]{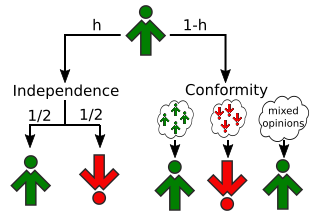}\hspace{3cm}
\includegraphics[scale=0.35]{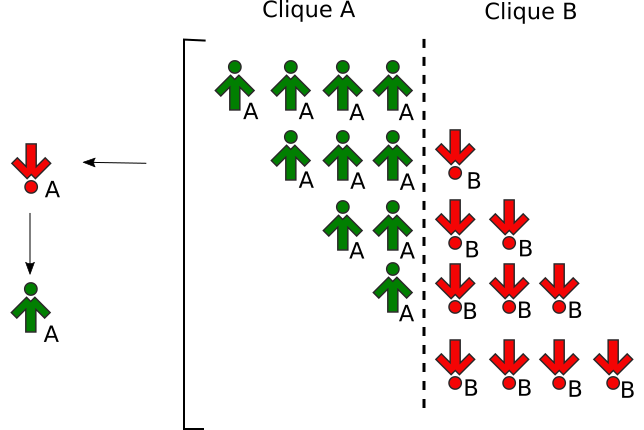}
\caption{Left plot: a schematic representation of the opinion update of a single agent being initially in the up state. Right plot: possible choices of the influence group in case $q=4$ perceived as unanimous by the target agent.\label{fig:model}}
\end{figure}

We will introduce the independence to the model in the situation-oriented manner~\cite{SZN14b,JED17}.
In a given time step, a target of influence will behave independently with probability $h$ or will become a conformist with probability $1-h$ (left plot in Fig.~\ref{fig:model}). Thus, an additional control parameter $h$ will be used to simulate the impact of the situational factors on the behavior of agents. Within this approach every agent may change his behavior from step to step and sometimes act independently, sometimes like a conformist.

\subsection{Quenched and annealed disorder models}
\label{sec:quenched vs annealed}

\begin{figure}
\centering
\includegraphics[scale=0.5]{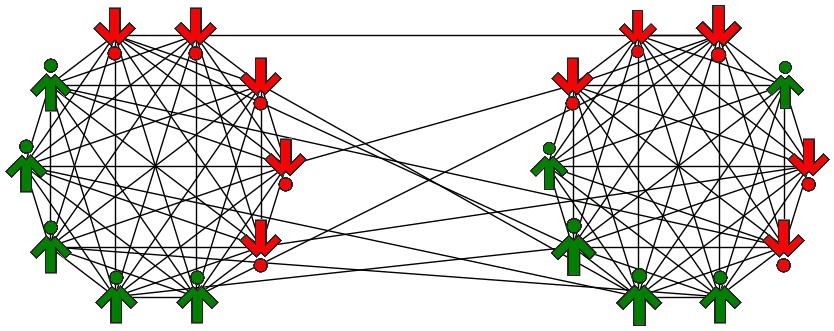}\ 
\caption{The quenched disorder model: a double-clique network consists of two complete graphs (cliques) connected with $LN^2$ cross links with each other. Here, $N$ is the size of each clique, $N^2$ is the maximal number of the cross-links and $L$ the fraction of the existing ones.\label{fig:two_clique}}
\end{figure}

In Ref.~\cite{KRU17}, two versions of the model were considered. In the quenched disorder one, two cliques of size $N$ each are connected with $L\times N^2$ number of cross-links. The parameter $L$ is simply the fraction of the existing cross-links, $N^2$ - their maximum number. Once the links between cliques are chosen randomly they remain fixed - the resulting network does not change in time during the evolution of the system (Fig.~\ref{fig:two_clique}).

\begin{figure}
\centering
\includegraphics[scale=0.5]{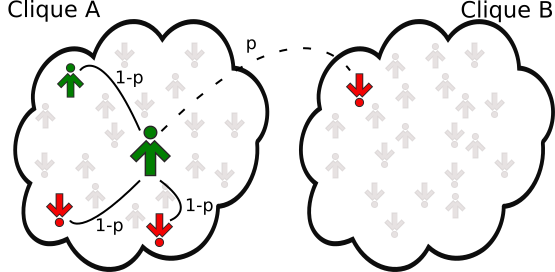}
\caption{The annealed disorder model: every agent from one clique is connected with probability $p$ with an agent from the other clique, and with probability $1-p$ with an agent from its own clique. See Eq.~(\ref{eq:p}) for the relationship between the probability $p$ and the fraction of cross links $L$. \label{fig:p}}
\end{figure}

In the annealed version of the model we introduced the probability $p$ of choosing one cross-link out of all edges in the double-clique network,
\begin{equation}
p=\frac{LN^2}{LN^2+2\frac{N(N-1)}{2}}\simeq \frac{L}{L+1}.
\label{eq:p}
\end{equation}
Instead of working with the fixed-cross links, we assume that every agent from one clique is connected with probability $p$ with an agent from the other clique, and with probability $1-p$ with an agent from its own clique (Fig.~\ref{fig:p}). Technically, this approximation is nothing but an average of the quenched disorder model over different configurations of cross-links in the network.

If the number of cross-links is smaller than their maximum number, the agents in the quenched disorder model differ from each other, because some of them may have no connections to the other clique, some others - multiple ones. While it can be handled with ease within a computer simulation, this feature constitutes usually a challenge for mathematical modeling due to the necessity to perform a quenched average over the disorder~\cite{LIU05}. The annealed model is easier in the sense that it allows for mathematical treatment.

\subsection{Updating rules of the models}

To recap, we consider a set of $2N$ agents, each of whom may be in one of two possible states reflecting an opinion on some given issue: $S_i=-1$ or $S_i=1$ for $i=1,2,\dots,2N$. We put the agents on a double-clique network, which consists of two complete graphs of $N$ nodes connected with $L\times N^2$ cross links.

We assume that the type of social response of agents depends on their group identity. Thus, an agent will strive for agreement within his/her own clique (conformity) and simultaneously will challenge the opinions of individuals from the other clique (anticonformity). As in Ref.~\cite{KRU17}, we introduce the notion of signals to the $q$-voter model and slightly alter the concept of unanimity of the influence group in order to account for the fact that an agent may act as both conformist and anticonformist at the same time. A signal is just a state of the neighbor when coming from the target’s clique or its inverted state otherwise. The target of influence changes its opinion only if all members of the influence group emit the same signal (see right plot in Fig.~\ref{fig:model}).

We will use Monte Carlo simulations with a random sequential updating scheme as the main tool to analyze the models. Each Monte Carlo step in a simulations consists of $2\times N$ elementary events, each of which may be divided into the following substeps with $\Delta t = \frac{1}{2N}$:
\begin{enumerate}
\item Pick a target agent at random (uniformly from $2N$ nodes). \label{step:pick}
\item Draw a random number form a uniform distribution, $r\sim U(0,1)$.
\item If $r<h$ (i.e. with probability $h$), the agent is independent:
    \begin{enumerate}
        \item Change its state with probability $1/2$, i.e. $S_i(t+\Delta t)=-S_i(t)$.
        \item Go to step~\ref{step:pick}.
    \end{enumerate}
\item If $r>h$ (i.e. with probability $1-h$), the agent is subject to social influence:
    \begin{enumerate}
        \item Choose randomly a group of $q$ distinct neighbors of the target node:
        \begin{description}
            \item[Quenched model] simply look at the actual neighbors of the target (sampling with replacement). 
            \item[Annealed model] first decide to which clique every member of the influence group will belong (with probability $1-p$ to the target’s clique, with $p$ to the other one), then choose the member randomly from the appropriate clique (Fig.~\ref{fig:p}). 
        \end{description}
        \item Convert the states of the group members to signals. Assume that the signals of the neighbors from target’s clique are equal to their states. Invert the states when from the other clique.
        \item Calculate the total signal of the influence group by summing up individual signals of its members.
        \item If the total signal is equal to $\pm q$ (i.e. all group members emit the same signal), the target changes its opinion accordingly (right plot in Fig.~\ref{fig:model}). Otherwise nothing happens.
    \end{enumerate}
\item Go to step~\ref{step:pick}.
\end{enumerate}

\section{Mathematical analysis of the annealed model}
\subsection{Quantities of interest}

The macroscopic state of an opinion dynamics model is usually described by either the concentration of agents in state $+1$ or the average opinion (i.e. magnetization in physical systems). Having in mind, that the total number of agents in our model is $2N$, we get the following formula for the concentration:
\begin{equation}
c(t)=\frac{N^{\uparrow}(t)}{2N}.
\end{equation}
Here, $N^{\uparrow}(t)$ stands for the number of agents in state $+1$. Similarly, the average opinion is given by 
\begin{equation}
m(t) =\frac{1}{2N}\sum_{i=1}^{2N}S_i =\frac{N^{\uparrow}(t)-N^{\downarrow}(t)}{2N},
\end{equation}
where $N^{\downarrow}(t)$ denotes the number of agents in state $-1$. Both quantities may be used interchangeably, because there is a simple relation between them: 
\begin{equation}
m(t)=2c(t)-1.
\end{equation}

Knowing the concentration of the entire system may be not enough to describe it uniquely in case of the double-clique topology. For instance, the value $c(t)=1/2$ may correspond to no ordering in the system (i.e. a perfect mixture of $+1$ and $-1$ states in both cliques) or to polarization (all agents in state $+1$ in one clique and in state $-1$ in the other). That is why it will be more insightful to calculate the above quantities for single cliques rather than for the entire system,
\begin{eqnarray}
c_X (t) & = & \frac{N_X^{\uparrow}(t)}{N},~~X=A,B,\\
m_X(t) & = & \frac{1}{N}\sum_{i=1}^{N}S_{X,i} =\frac{N_X^{\uparrow}(t)-N_X^{\downarrow}(t)}{2}.\nonumber
\end{eqnarray}
The interpretation of their values is summarized in Table~\ref{tab:quantities}.

\begin{table}
\centering
\begin{tabular}{p{5cm}|c|c}
\hline \hline
Meaning & $c_X(t)$ & $m_X(t)$ \\ 
\hline 
Positive consensus (all agents in clique $X$ in state $+1$)& $c_X=1$ & $m_X=1$ \\ 
\hline 
Partial positive consensus (majority of agents in clique $X$ in state $+1$) & $1/2 < c_X<1$ & $0 < m_X<1$ \\ 
\hline 
No ordering in clique $X$ & $c_X=1/2$ & $m_X=0$ \\ 
\hline 
Partial negative consensus (majority of agents in clique $X$ in state $-1$) & $0<c_X<1/2$ & $-1 < m_X<0$ \\ 
\hline 
Negative consensus (all agents in clique $X$ in state $-1$) &  $c_X=0$ & $m_X=-1$ \\ 
\hline \hline
\end{tabular} 
\caption{Interpretation of different values of the concentration $c_X(t)$ and the average opinion $m_X(t)$ within a single clique $X$.\label{tab:quantities} }
\end{table}

\subsection{Transition probabilities}

The random sequential updating scheme in our model means that in every single time step $\Delta t =1/2N$ only one agent can change its opinion. Three scenarios are possible: (1) the total amount of agents in state $+1$ in a clique may increase  by 1, (2) it may decrease by $1$ or (3) it may remain unchanged. 

Let us have a look at the first of the above scenarios. The number of agents in state $+1$ in one clique - say A - can increase by 1 only if:
\begin{enumerate}
\item a target from clique $A$ is chosen (probability $1/2$),
\item the target is in state $-1$ (probability $1-c_A$),
\item it flips due to independence (probability $h/2$) or follows an influence group emitting signal $+q$.
\end{enumerate} 
Thus, the transition probability for such an event will be given by
\begin{equation}
\Pr \left\{ N_{A}^\uparrow\left( t+\Delta t\right) =N_{A}^\uparrow\left( t\right)
+1\right\} =\frac{1}{2}\left( 1-c_{A}(t) \right) \left( \frac{1}{2}h+\left(1-h\right)\left[ \left(
1-p\right) c_{A}\left( t\right) +p\left( 1-c_{B}\left( t\right) \right)
\right]^{q}\right). \label{eq:transprob1}
\end{equation}
One can easily check that the term of the form $(u+v)^q$ in the above equation is just the generating function for the probabilities of those compositions of $q$ members of an influence group that can cause an opinion switch event (right plot in Fig.~\ref{fig:model}). Similarly, the number of agents in state $+1$ in clique $A$ decreases by 1 if:
\begin{enumerate}
\item a target from clique $A$ is chosen (probability $1/2$),
\item the target is in state $+1$ (probability $c_A$),
\item it flips due to independence (probability $h/2$) or follows an influence group emitting signal $-q$.
\end{enumerate} 
These conditions lead to the following transition probability:
\begin{equation}
\Pr \left\{ N_{A}^\uparrow\left( t+\Delta t\right) =N_{A}^\uparrow\left( t\right)
-1\right\} = \frac{1}{2}c_{A}\left( t\right) \left( \frac{1}{2}h+\left(1-h\right)\left[ \left( 1-p\right)
\left( 1-c_{A}\left( t\right) \right) +pc_{B}\left( t\right) \right]^{q}\right). \label{eq:transprob2}
\end{equation}
It is also possible that the number of agents in state $+1$ remains unchanged in elementary time step. The probability of this event is simply given by:
\begin{equation}
\Pr \left\{ N_{A}^\uparrow\left( t+\Delta t\right) =N_{A}^\uparrow\left( t\right) \right\}
= 1-\Pr \left\{ N_{A}^\uparrow\left( t+\Delta t\right) =N_{A}^\uparrow\left( t\right)
+1\right\} -\Pr \left\{ N_{A}^\uparrow\left( t+\Delta t\right) =N_{A}^\uparrow\left(
t\right) -1\right\}. \label{eq:transprob3}
\end{equation}
Analogous considerations for clique B yield:
\begin{eqnarray}
\Pr \left\{ N_{B}^\uparrow\left( t+\Delta t\right) =N_{B}^\uparrow\left( t\right)
+1\right\} &=&\frac{1}{2}\left( 1-c_{B}\left( t\right) \right) \left( \frac{1}{2}h+\left(1-h\right)\left[ \left(
1-p\right) c_{B}\left( t\right) +p\left( 1-c_{A}\left( t\right) \right)
\right] ^{q} \right),\nonumber\\
\Pr \left\{ N_{B}^\uparrow\left( t+\Delta t\right) =N_{B}^\uparrow\left( t\right)
-1\right\} &=&\frac{1}{2}c_{B}\left( t\right) \left( \frac{1}{2}h+\left(1-h\right)\left[ \left( 1-p\right)
\left( 1-c_{B}\left( t\right) \right) +pc_{A}\left( t\right) \right] ^{q}\right), \label{eq:transprob4}\\
\Pr \left\{ N_{B}^\uparrow\left( t+\Delta t\right) =N_{B}^\uparrow\left( t\right) \right\}
&=&1-\Pr \left\{ N_{B}^\uparrow\left( t+\Delta t\right) =N_{B}^\uparrow\left( t\right)
+1\right\} -\Pr \left\{ N_{B}^\uparrow\left( t+\Delta t\right) =N_{B}^\uparrow\left(
t\right) -1\right\}.\nonumber
\end{eqnarray}

\subsection{Expectation values}

Given the states of the cliques at time $t$, we can derive the expectation values for the numbers of agents in state $+1$ at time $t+\Delta t$ from the above expressions::
\begin{eqnarray}
E\left( N_{A}^\uparrow\left( t+\Delta t\right) \right) = N_{A}^\uparrow\left( t\right) &+&
\frac{1}{2}\left( 1-c_{A}\left( t\right) \right) \left( \frac{1}{2}h+\bar{h}\left[ \bar{p}c_{A}\left(
t\right) +p\left( 1-c_{B}\left( t\right) \right) \right] ^{q}\right)\nonumber\\
&-&\frac{1}{2}
c_{A}\left( t\right) \left( \frac{1}{2}h+\bar{h}\left[ \bar{p}\left( 1-c_{A}\left( t\right) \right)
+pc_{B}\left( t\right) \right] ^{q}\right), \nonumber \\
E\left( N_{B}^\uparrow\left( t+\Delta t\right) \right) = N_{B}^\uparrow\left( t\right) &+&
\frac{1}{2}\left( 1-c_{B}\left( t\right) \right) \left( \frac{1}{2}h+\bar{h}\left[ \bar{p}c_{B}\left(
t\right) +p\left( 1-c_{A}\left( t\right) \right) \right] ^{q}\right)\nonumber\\
&-&\frac{1}{2}
c_{B}\left( t\right) \left( \frac{1}{2}h+\bar{h}\left[ \bar{p}\left( 1-c_{B}\left( t\right) \right)
+pc_{A}\left( t\right) \right] ^{q}\right).\label{eq:expect}
\end{eqnarray}
The abbreviations $\bar{p}=1-p$ and $\bar{h}=1-h$ were used here for the sake of readability.

\subsection{Asymptotic dynamical system}
We would like to derive  a limiting dynamical system for $N\rightarrow \infty$ from Eqs.~(\ref{eq:expect}). Let us first divide the above equations by $N$:
\begin{eqnarray}
E\left( c_{A}\left( t+\Delta t\right) \right) - c_{A}\left( t\right) &=&
\frac{1}{2N}\left( 1-c_{A}\left( t\right) \right) \left( \frac{1}{2}h+\bar{h}\left[ \bar{p}c_{A}\left(
t\right) +p\left( 1-c_{B}\left( t\right) \right) \right] ^{q}\right)\nonumber\\
& &-\frac{1}{2N}
c_{A}\left( t\right) \left( \frac{1}{2}h+\bar{h}\left[ \bar{p}\left( 1-c_{A}\left( t\right) \right)
+pc_{B}\left( t\right) \right] ^{q}\right), \nonumber \\
E\left( c_{B}\left( t+\Delta t\right) \right) - c_{B}\left( t\right) &=&
\frac{1}{2N}\left( 1-c_{B}\left( t\right) \right) \left( \frac{1}{2}h+\bar{h}\left[ \bar{p}c_{B}\left(
t\right) +p\left( 1-c_{A}\left( t\right) \right) \right] ^{q}\right)\nonumber\\
& &-\frac{1}{2N}
c_{B}\left( t\right) \left( \frac{1}{2}h+\bar{h}\left[ \bar{p}\left( 1-c_{B}\left( t\right) \right)
+pc_{A}\left( t\right) \right] ^{q}\right).\label{eq:expect2}
\end{eqnarray}
It is very likely that in the limit $N\rightarrow\infty$ the random variables $c_X = \frac{N^\uparrow_X}{N}$ localize and hence become almost surely equal to their expectations. Hence we get:
\begin{eqnarray}
\frac{c_{A}\left( t+\Delta _{N}\right) - c_{A}\left( t\right)}{\Delta t} &=&  
\left( 1-c_{A}\left( t\right) \right) \left( \frac{1}{2}h+\bar{h}\left[ \bar{p}c_{A}\left(
t\right) +p\left( 1-c_{B}\left( t\right) \right) \right] ^{q}\right)\nonumber\\
& &-
c_{A}\left( t\right) \left( \frac{1}{2}h+\bar{h}\left[ \bar{p}\left( 1-c_{A}\left( t\right) \right)
+pc_{B}\left( t\right) \right] ^{q}\right), \nonumber \\
\frac{c_{B}\left( t+\Delta t\right) - c_{B}\left( t\right)}{\Delta t} &=& 
\left( 1-c_{B}\left( t\right) \right) \left( \frac{1}{2}h+\bar{h}\left[ \bar{p}c_{B}\left(
t\right) +p\left( 1-c_{A}\left( t\right) \right) \right] ^{q}\right)\nonumber\\
& &-
c_{B}\left( t\right) \left( \frac{1}{2}h+\bar{h}\left[ \bar{p}\left( 1-c_{B}\left( t\right) \right)
+pc_{A}\left( t\right) \right] ^{q}\right).\label{eq:expect3}
\end{eqnarray}
Taking the limit $N\rightarrow\infty$ and denoting the limiting variables $c_A$ and $c_B$ by $x$ and $y$ we arrive at:
\begin{eqnarray}
\frac{{\rm d} x}{{\rm d} t} &=& \left( 1-x \right) \left( \frac{1}{2}h+\bar{h}\left[ \bar{p}x +p\left( 1-y \right) \right] ^{q}\right)
-x \left( \frac{1}{2}h+\bar{h}\left[ \bar{p}\left( 1-x \right)+py \right] ^{q}\right), \nonumber \\
\frac{{\rm d} y}{{\rm d} t} &=& \left( 1-y \right) \left( \frac{1}{2}h+\bar{h}\left[ \bar{p}y +p\left( 1-x \right) \right] ^{q}\right)
-y \left( \frac{1}{2}h+\bar{h}\left[ \bar{p}\left( 1-y \right)+px \right] ^{q}\right).\label{eq:dynamical}
\end{eqnarray}

\subsection{Direction fields and stationary points}

\begin{figure}[tbp]
	\centering
	\includegraphics[width=.40\textwidth]{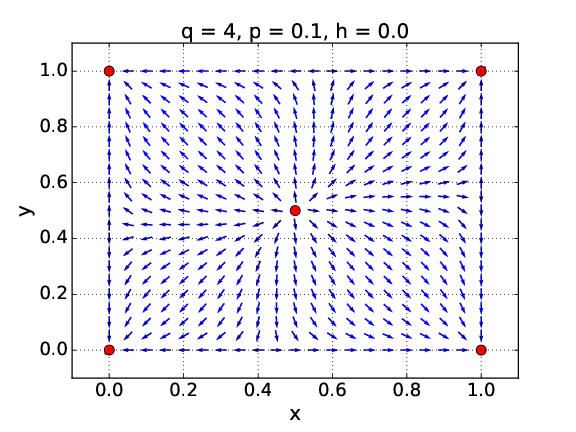}
	\includegraphics[width=.40\textwidth]{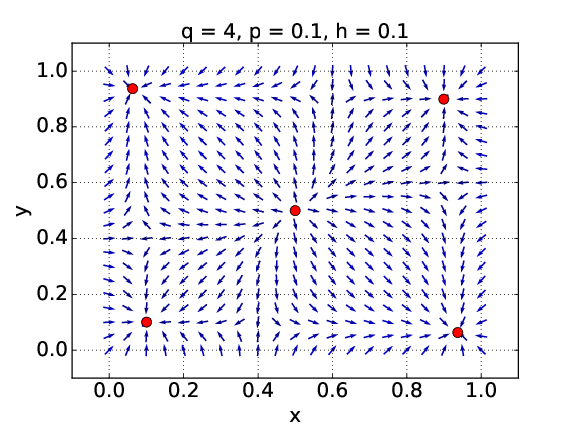}
	\includegraphics[width=.40\textwidth]{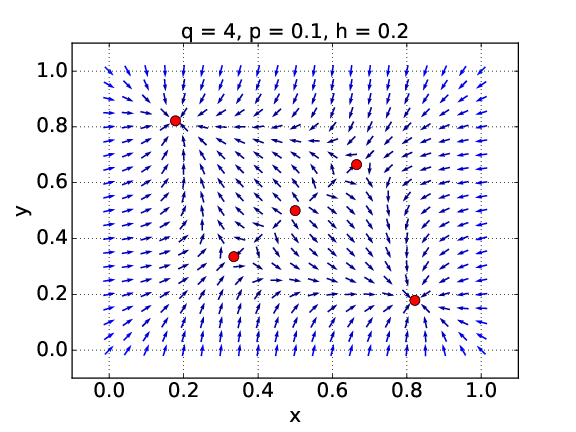}
	\includegraphics[width=.40\textwidth]{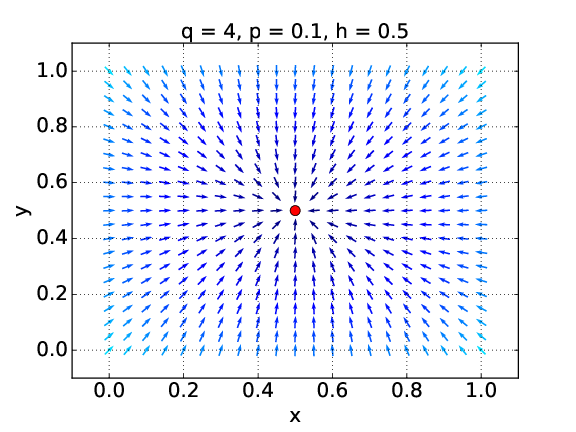}
	\includegraphics[width=.40\textwidth]{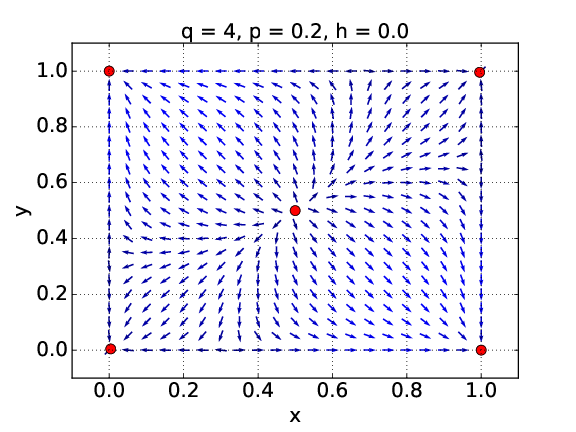}
	\includegraphics[width=.40\textwidth]{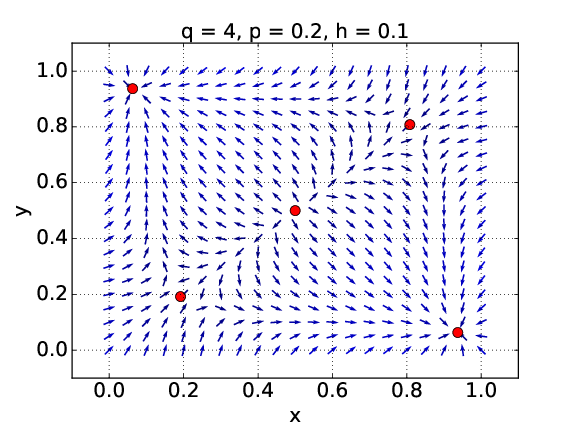}
	\includegraphics[width=.40\textwidth]{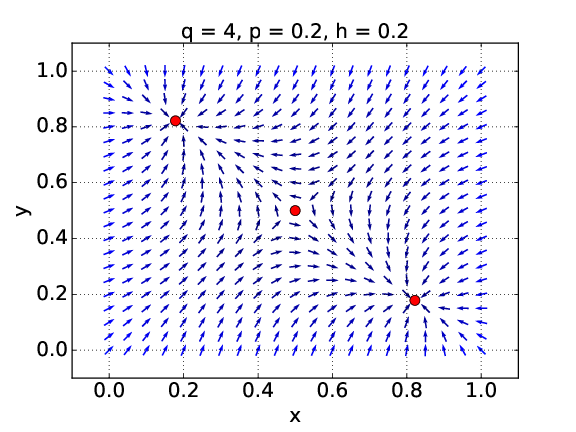}
    \includegraphics[width=.40\textwidth]{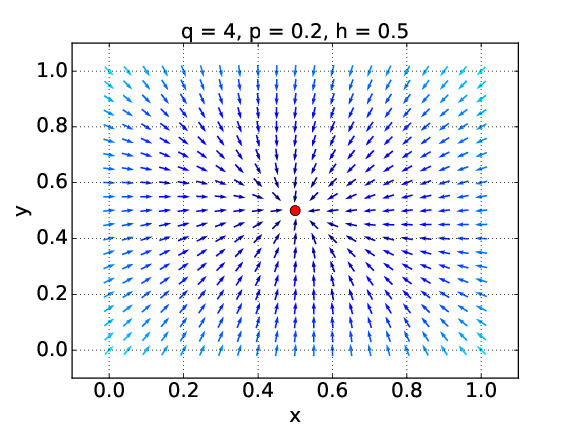}

	\caption{The annealed model: direction fields of the model described by Eq.~(\ref{eq:dynamical}) with fixed points marked with circles for different values of independence $h$ and two values of parameter $p$, $0.1$ (left column) and $0.2$ (right column)}
	\label{fig:anfields}
\end{figure}

The set of equations~(\ref{eq:dynamical}) is too cumbersome to solve it analytically. However, we still can generate direction fields of the set in order to graphically trace out solution curves for different initial values~\cite{STR94}. Results for different independence levels $h$ and two different probabilities of an inter-clique connection $p$ are shown in Fig.~\ref{fig:anfields}: the left column contains the plots for $p=0.1$, the right one corresponds to $p=0.2$. The values of $h$ are equal to $0.0$, $0.1$, $0.2$ and $0.5$ (from top to bottom).  Note that the case $h=0$ is nothing but our original model with no independence, extensively studied in Ref.~\cite{KRU17}.

From the flows in the state plane it follows that for $p=0.1$ and $h=0$ there are five stationary points (already marked with dots in the plots). Two attractors, $P_1=(0,1)$ and $P_2=(1,0)$, correspond to a polarized state of the system, i.e. all agents in one clique are in state $+1$ and in the other - in state $-1$. There are two other symmetric attractors, $C_1$ and $C_2$, very close to the coordinates $(0,0)$ and $(1,1)$. Thus the state of (almost) complete consensus is possible in the system as well. The remaining point $R$ is a repeller, because the system tends to evolve away from it. 

To find the exact coordinates of the stationary points, we just set $x^\prime$ and $y^\prime$ equal to zero in Eq.~(\ref{eq:dynamical}) and solve the resulting set of equations with respect to $x$ and $y$,
\begin{eqnarray}
0 &=& \left( 1-x \right) \left( \frac{1}{2}h+\bar{h}\left[ \bar{p}x +p\left( 1-y \right) \right] ^{q}\right)
-x \left( \frac{1}{2}h+\bar{h}\left[ \bar{p}\left( 1-x \right)+py \right] ^{q}\right), \nonumber \\
0 &=& \left( 1-y \right) \left( \frac{1}{2}h+\bar{h}\left[ \bar{p}y +p\left( 1-x \right) \right] ^{q}\right)
-y \left( \frac{1}{2}h+\bar{h}\left[ \bar{p}\left( 1-y \right)+px \right] ^{q}\right).\label{eq:equilibria}
\end{eqnarray}
For $p=0.1$ and $h=0.0$ we obtain:
\begin{eqnarray}
&& P_1 = (0,1),~~~~ P_2=(1,0),\\
&& C_1 = (0.00015,0.00015),~~~~ C_2=(0.99985,0.99985),\nonumber\\
&& R = (0.5,0.5).\nonumber 
\end{eqnarray}

Introducing a small level of independence ($h=0.1$ and $0.2$) into the model does not change the classification of the stationary points for $p=0.1$. However, they are now shifted towards the center of the state plane meaning that the complete polarization and (almost) complete consensus have changed to the partial ones. Although these states are still characterized by a majority of agents sharing the same opinion, due to the fluctuations induced by independence there is now always a minority having the opposite opinion. At a high independence level ($h=0.5$) the point $R=(0.5,0.5)$ becomes an attractor and the other stationary points disappear. 

The situation for $p=0.2$ is similar, however now we explicitly see what happens with the system between the state with five fixed points and the state with only a single one. For $h=0.2$, the consensus attractors $C_1$ and $C_2$ have already disappeared. The polarization ones are still there, but they are closer to the center of the plane. And the repeller $R=(0.5,0.5)$ becomes hyperbolic. With further increase of $h$ the polarization attractors will disappear as well and the point $R$ will become an attractor (see case $h=0.5$). 

Compared to the model without independence~\cite{SIE16,KRU17} we observe now an additional dynamical phase transition in the system - for independence levels large enough it is entering the disordered phase with the vanishing magnetization in every clique as the asymptotic state. 

\section{Numerical results}
\label{sec:Results}

As already pointed in Sec.~\ref{sec:quenched vs annealed}, the quenched version of the model does not allow for a mathematical treatment similar to the one presented in the previous section. That is why we will resort to Monte Carlo simulations of the model and compare them with the numerical solutions of the Eq.~(\ref{eq:dynamical}) for the annealed one.

We will assume that the number of agents in every clique in the quenched model is $N=100$. Although the size of the system may seem to be very small, it was already shown in Refs.~\cite{SIE16,KRU17} that increasing the size does not qualitatively change the outcome of the simulations, but it takes substantially longer to finish them.

We considered in our analysis influence groups of sizes ranging from 2 to 6, with the upper bound motivated with the conformity experiments by Asch~\cite{ASC55}. Qualitatively, the results turned out to be independent of the actual value of $q$. Thus, we decided to present the results for $q=4$, a value often used in the analysis of the $q$-voter model and its extensions. 

If not stated otherwise, the results of the simulations were averaged over 1000 independent runs. In most of the cases the asymtotic state was reached quickly, in less than 500 Monte Carlo steps. We used our own codes written in C++, Python and Matlab. 

As for the initial condition, we took the total positive consensus, i.e. all agents being in the state $+1$. As already pointed out in Ref.~\cite{SIE16}, this choice may be treated as a result of the following scenario. Two cliques with a natural tendency to disagree with each other evolve at first independently. They get in touch by chance and establish some cross-links to the other group after they both reached consensus on a given issue.

\subsection{Time evolution of the system}

\begin{figure}[tbp]
	\centering
	\includegraphics[width=.30\textwidth]{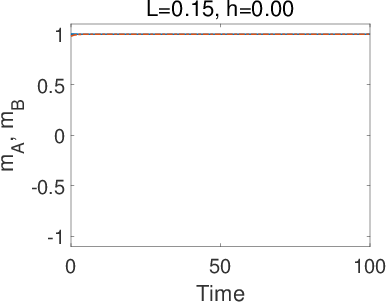}
	\includegraphics[width=.30\textwidth]{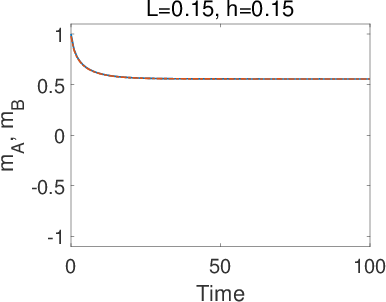}
	\includegraphics[width=.30\textwidth]{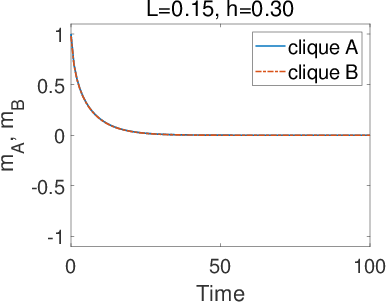}\vspace{1cm}\\
	\includegraphics[width=.30\textwidth]{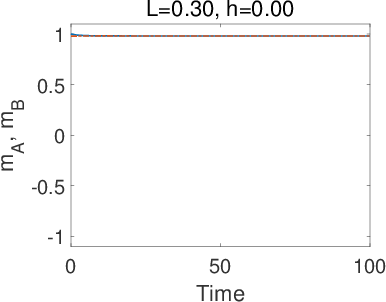}
	\includegraphics[width=.30\textwidth]{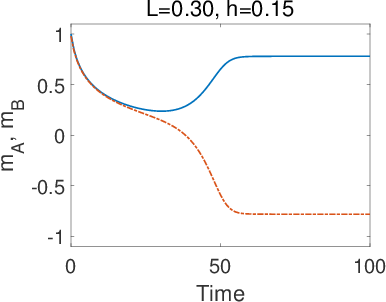}
	\includegraphics[width=.30\textwidth]{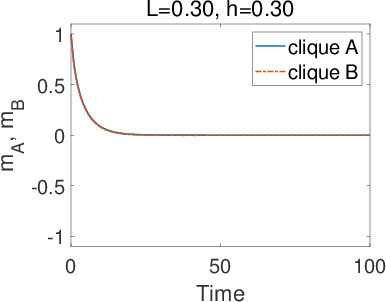}\vspace{1cm}\\
	\includegraphics[width=.30\textwidth]{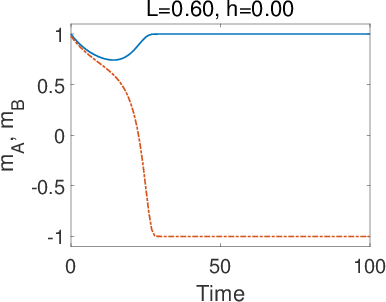}
	\includegraphics[width=.30\textwidth]{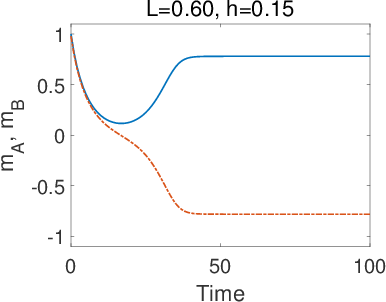}
	\includegraphics[width=.30\textwidth]{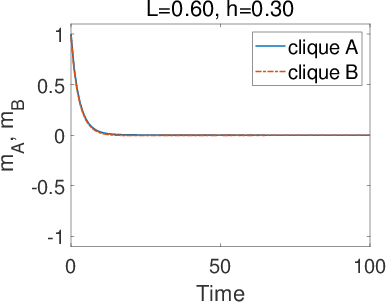}
	\caption{(Color online) Magnetizations as a function of time in the annealed model for different values of $L$ and $h$.}
	\label{fig:aevol}
\end{figure}

The asymptotic dynamical system for the annealed model, given by Eq.~(\ref{eq:dynamical}), was solved numerically. Results for different values of parameters $p$ and $h$ are shown in Fig.~\ref{fig:aevol}.
As already concluded from the direction fields (Fig.~\ref{fig:anfields}), in the absence of independence (left column in Fig.~\ref{fig:aevol}) consensus in both cliques is observed for a small number of cross-links. More connections between the cliques drive the system to a polarized state. The picture is different already for a moderate level of independence in the model (middle column in Fig.~\ref{fig:aevol}). We still observe consensus if the cliques are poorly connected. However, polarization sets in  at much lower number of cross links. Moreover, both the consensus and polarization are partial, because due to independence there is always a group of agents not going with the majority. Increasing the independence level destroys the ordering in the system and the model ends up in the asymptotic state with no magnetization. This last result is independent on the number of cross-links between the cliques. 

\begin{figure}[tbp]
	\centering
	\includegraphics[width=.30\textwidth]{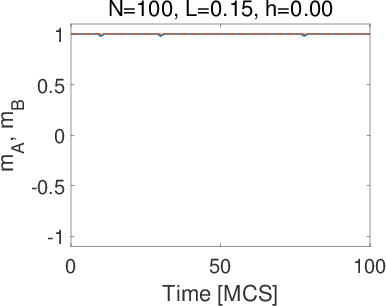}
	\includegraphics[width=.30\textwidth]{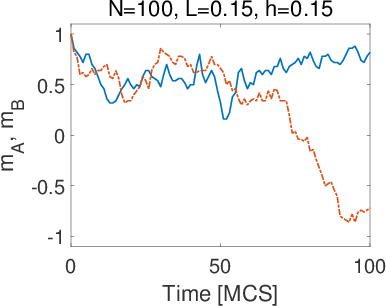}
	\includegraphics[width=.30\textwidth]{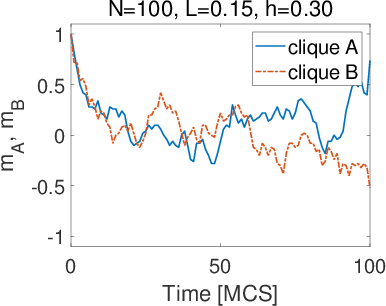}\vspace{1cm}\\
	\includegraphics[width=.30\textwidth]{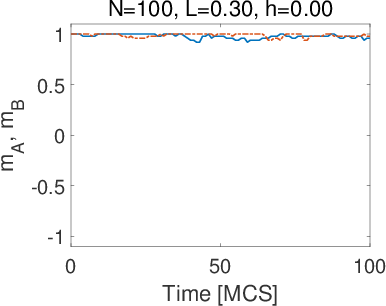}
	\includegraphics[width=.30\textwidth]{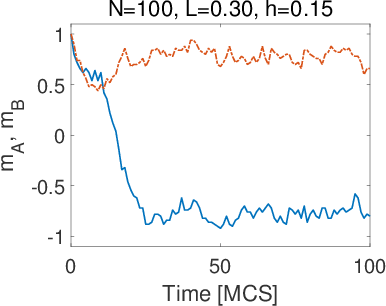}
	\includegraphics[width=.30\textwidth]{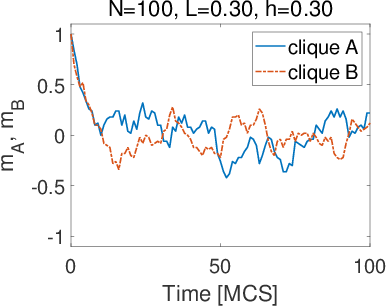}\vspace{1cm}\\
	\includegraphics[width=.30\textwidth]{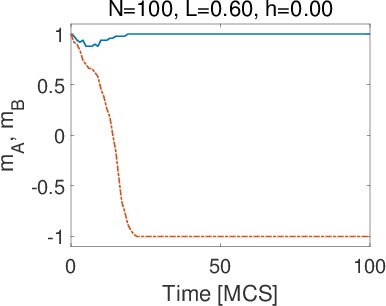}
	\includegraphics[width=.30\textwidth]{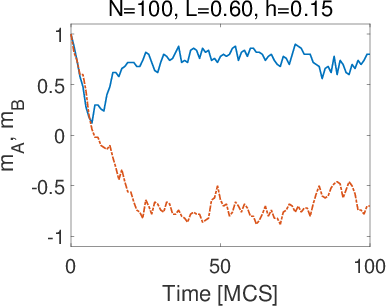}
	\includegraphics[width=.30\textwidth]{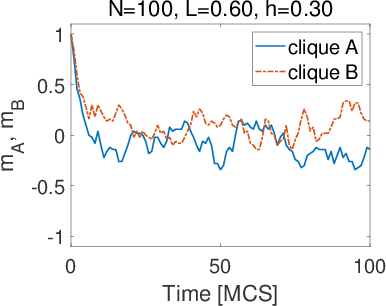}
	\caption{(Color online) Magnetizations as a function of time in the quenched model for different values of $L$ and $h$. All trajectories shown in the plots are from single runs.}
	\label{fig:qevol}
\end{figure}

Monte Carlo simulations of the quenched version of the model produce similar output (see Fig.~\ref{fig:qevol}).
Despite the fluctuations due to the stochastic nature of the dynamics, the curves are qualitatively the same for most of the parameter groupings. The only exception is the case $(L,h)=(0.15,0.15)$, in which we get polarization and not partial consensuns as in the annealed model. It indicates that the critical value of $L$ for the dynamical consensus-polarization phase transition is smaller for the quenched model, in agreement with our previous findings for the models without independence~\cite{SIE16,KRU17}.

\subsection{Influence of independence on the system}
\label{ssec:influence}

All results up to this point already suggest that there are already three effects resulting from introduction of independence into the models: (1) final concentrations of agents sharing the same opinion are diminished, (2) the critical values of $L$ at the consensus-polarization transition are smaller and (3) an additional dynamical phase transition from the polarized state to an disordered one occurs in the system. 

\begin{figure}[tbp]
	\centering
	\includegraphics[width=.49\textwidth]{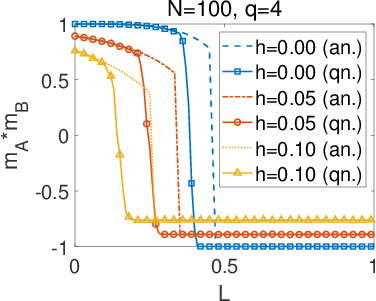}
    \caption{(Color online) Comparison between two models: averaged product of magnetizations of the cliques as a function of parameter $L=\frac{p}{1-p}$ for $N=100$, $q=4$ and three values of independence $h=0,0.05,0.1$.}
	\label{fig:anmag}
\end{figure}

To elaborate on those findings, let us first have a look at Fig.~\ref{fig:anmag}. In this plot, the product of final magnetizations in the cliques is presented as the function of the fraction of cross-links $L$ at different levels of independence. The case $h=0$ (no independence) corresponds to the original models from Refs.~\cite{SIE16,KRU17}. All calculations were done with the total positive consensus as the initial condition. We see that for values smaller than a critical value both cliques always end up in consensus. In other words, in this regime the intra-clique conformity wins with the inter-clique anticonformity and both communities are able to maintain their initial consensus, at least partially. 
Larger values of $L$ are needed for the negative ties to take over and to push the system into a polarized state. The impact of independence is two-fold. First, the final magnetizations have been pushed away from the values $\pm 1$ even in the case $L=0$, i.e. the total consensus have changed to a partial one. Since it corresponds to the weakening of the force exerted by conformity, one would expect than in this case less cross-links between the cliques are needed to polarize the system. Indeed, as we see in Fig.~\ref{fig:anmag}, the critical value of $L$ shifts to the left with increasing independence $h$. The actual critical values of $L$ in both models for several values of $h$ are listed in Table~\ref{tab1}.
\begin{table}
\centering
\begin{tabular}{c|c|c}
\hline\hline
independence $h$ & $L$ (quenched) & $L$ (annealed)\\
\hline
0.00 & 0.41 & 0.46\\
0.05 & 0.29 & 0.35\\
0.10 & 0.19 & 0.26\\
0.15 & 0.11 & 0.18\\
0.20 & 0.01 & 0.11\\
0.25 & 0.01 & 0.04\\
\hline \hline
\end{tabular}
\caption{Critical values of $L$ for different independence levels $h$.\label{tab1}}
\end{table}%

It should be noted that for each value of $h$, there is a difference in the critical values between the quenched and annealed models. It is mainly a consequence of different system sizes - while Eq.~(\ref{eq:dynamical}) defining the annealed model was derived for an infinite system, we used only 200 agents in the simulations of the quenched one.  It has been shown in Ref.~\cite{KRU17} that the discrepancy between the models decreases with the increasing size of the simulated system. We expect the models to converge for $N\rightarrow\infty$ despite the subtle changes in their dynamics.

\begin{figure}[tbp]
	\centering
	\includegraphics[width=.45\textwidth]{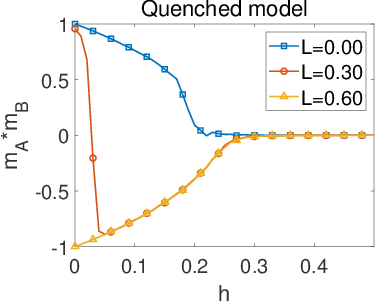} \hfill
    \includegraphics[width=.45\textwidth]{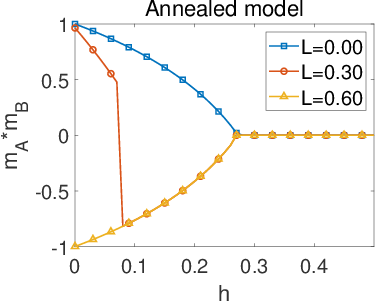}
    \caption{(Color online) Averaged product of magnetization of the cliques as a function of independence $h$ for $N=100$, $q=4$ and three values of parameter $L=0,0.3,0.6$ in the quenched model (left) and in the annealed (right).}
	\label{fig:anmagh}
\end{figure}

To complete the picture, let us investigate how the product of magnetizations changes with independence $h$ (Fig.~\ref{fig:anmagh}). At $L=0$ (no connection between the cliques), independence continuously  destroys the ordering in both communities. Finally, above a critical value $h^*$, the system enters the disordered phase with no magnetization in the cliques. The phase portrait at $L=0.30$ (moderate number of connections) is more interesting. For small values of $h$ the system maintains the partial consensus, then we observe a transition to the polarized state. The magnetizations in the now antagonistic cliques are diminishing with further increasing of $h$. Finally, the system reaches the disordered phase. At $L=0.60$, the system is already polarized, even for $h=0$. Increasing $h$ introduces disorder into the cliques. Again, there is no ordering above the critical value of $h$. 

As already discussed earlier in this section, there are some differences in absolute values between the annealed and quenched models, but the picture for the annealed case is qualitatively the same (right plot in Fig.~\ref{fig:anmagh}).

It is worth to note that the critical value of $h$ for the polarization-disorder transition does not depend on $L$.  
%
\section{Conclusions}
\label{sec:Conclusions}

The most important message from our previous study was that the consensus between two antagonistic communities is possible only if they are loosely connected with each other~\cite{SIE16,KRU17}. The more interactions between those communities take place the less probable it is that the entire system will share the same opinion. Instead, the anticonformity takes over and pushes the system to polarization. Those results were unexpected in the sense that they for instance supported the idea of the often criticized filter bubbles on social media~\cite{BAI18,PAR11}. Since those bubbles separate users from information that disagrees with their viewpoints, they may help to weaken the problem with polarization. However, the models we considered were very simple. For instance, they were lacking some typical answers to social influence~\cite{NAI00}.

In order to make the models more realistic, in this work we added independence as a response to social influence. From our results it follows that this additional manifestation of social interactions impacts the dynamics of the system in at least two ways. Small independence levels help anticonformity to take over and polarize the society. More technically speaking, they lower the critical ratio of cross-links between cliques needed to arrive at a polarized state. High independence levels destroy any ordering in the system. Consequently, the opinions of agents are perfectly mixed across the cliques and neither consensus nor polarization are observed.

To sum up, low (but present) independence levels seem to enhance the polarization of societies. Thus, they counteract the effects of the filter bubbles, which at least within our models foster consensus across the cliques. At high levels all manifestations of the interplay between conformity and anticonformity are suppressed by the noise induced due to independence.

\bibliographystyle{elsarticle-num}
\bibliography{polarization}

\end{document}